\documentclass[preprint,12pt]{elsarticle}

\usepackage{graphicx}
\usepackage{amssymb}
\usepackage{amsmath}
\usepackage{nicefrac}
\usepackage{color}
\usepackage[utf8]{inputenc}
\usepackage{url}
\usepackage{hyperref}
\usepackage{comment}

\journal{ }

\begin{document}

\begin{frontmatter}


\title{The Sensitivity of Power System Expansion Models} 



\author[dlr]{Bruno~U.~Schyska\corref{bruno.schyska@dlr.de}}
\author[fias]{Alexander~Kies}
\author[fias]{Markus~Schlott}
\author[dlr, uniol]{Lueder~von~Bremen}
\author[dlr]{Wided~Medjroubi}

\address[dlr]{DLR Institute of Networked Energy Systems, Oldenburg, Germany}
\address[fias]{Frankfurt Institute of Advanced Studies, Goethe University, Frankfurt, Germany}
\address[uniol]{ForWind Center for Wind Energy Research, Carl-von-Ossietzky-University, Oldenburg, Germany}

\begin{abstract}
Power system expansion models are a widely used tool for planning power systems, especially considering the integration of large shares of renewable resources. The backbone of these models is an optimization problem, which depends on a number of economic and technical parameters. Although these parameters contain significant uncertainties, the sensitivity of power system models to these uncertainties is barely investigated. 

In this work, we introduce a novel method to quantify the sensitivity of power system models to different model parameters based on measuring the additional cost arising from misallocating generation capacities. The value of this method is proven by three prominent test cases: the definition of capital cost, different weather periods and different spatial and temporal resolutions. We find that the model is most sensitive to the temporal resolution. Furthermore, we explain why the spatial resolution is of minor importance and why the underlying weather data should be chosen carefully.



\end{abstract}


\end{frontmatter}


\section{Introduction}\label{sec:introduction}

In order to address the issue of climate change and sustainability, energy systems around the world are undergoing major transitions.  In this context, large shares of weather-dependent power sources, such as solar photovoltaics (PV) and wind power, need to be integrated into existing systems. This is a challenging task, solutions proposed by literature are manifold \cite{lund2007renewable,connolly2010review}. They include the large-scale integration of storage technologies \cite{steinke2013grid,weitemeyer2016european}, the extension of the transmission grid \cite{rodriguez2014transmission, kies2016curtailment}, the overinstallation of renewable capacities \cite{heide2011reduced} or optimising the mix of different renewable generation sources, e.g. solar and hydro power \cite{ming2017optimizing}, wind power and solar PV \cite{jurasz2017integrating} or wind power and concentrated solar power \cite{santos2015combining}.

Options to modify the demand side have for instance been investigated by \citet{palensky2011demand} and \citet{zerrahn2015representation}. \citet{kies2016demand} found that demand-side management can balance generation-side fluctuations for a renewable share of up to 65\% in Europe. \citet{hirth2016system} and \citet{chattopadhyay2017impact} proposed to deploy \emph{system-friendly} wind turbines or PV modules, respectively, that are designed to resemble load patterns. Furthermore, vehicle-to-grid technologies have been proven to be useful for the system integration of renewables \cite{lund2008integration}. It is common understanding that the coupling of the sectors electricity, heat and transportation might reduce costs and hence support the power system transition. The synergies between these sectors have been investigated by \citet{brown2018synergies}.



Many of these solutions are the result of studies using so-called \emph{power system expansion models} (PSEM). Although most PSEM aim at finding optimal solutions for power system design, they may significantly vary in structure and in scope. For a list of models see for instance the Open Energy Platform (\url{https://openenergy-platform.org/factsheets/models/}).

Over the last years, PSEM have become progressively more complex. They include more and more aspects of power system in an increasing level of detail. Running these models is an increasingly challenging and expensive task. Consequently, a number of complexity reduction techniques has been proposed in the literature \cite{raventos2020evaluation}. They range from simple averaging and clustering techniques to reduce the temporal and spatial resolution to more complex methods to be able to model storage units with inconsistent time series. However, without being able to compute a representative reference, their skill with respect to \emph{real world examples} can hardly be quantified.

Furthermore, PSEM depend on a number of uncertain parameters. Assumptions made for costs or the availability of weather-dependent generation sources as well as the reduction of the model resolution to make it tractable introduce uncertainty \cite{trutnevyte2016does, mavromatidis2018uncertainty, schlachtberger2018cost}. This uncertainty clearly effects the interpretation of the simulation results. Recently, \citet{nacken2019integrated} applied a method called \emph{modelling to generate alternatives} (MGA) to a future German energy supply and showed that it produces a number of significantly different energy scenarios. MGA bases on changing the PSEM structure by setting the cost-optimal objective value plus some slack as a new constraint and exploring the resulting solution space \cite{brill1982modeling, price2017modelling, decarolis2016modelling}. Neumann und Brown \cite{neumann2019near} used a similar method to explore the near-optimal solution space of a cost-optimised European power system. They observed a high variance in the deployment of individual components near the optimal solution. Based on a \emph{global sensitivity analysis}, \citet{MORET2017597} found that the uncertainty of economic parameters has the highest influence on the results of an energy model. Similarly, \citet{RePEc:fae:ppaper:2019.04} investigated the robustness of a fully renewable power system model of France to uncertainties in future cost of generation technologies. They found that, although the optimal generation mix clearly depends on the respective cost for the different technologies, overall system costs are relatively insensitive. For an overview of methods applied in the context of uncertainty in power system modeling see \citet{YUE2018204}.

In addition, power infrastructure is planned decades in advance. During these decades, boundary conditions may considerably change due to climate change and climate variability. The effect of climate change on renewable power systems is a recent field of research. \citet{schlott2018impact} found that it is likely that climate change will increase the share of solar PV in a cost-optimal European power system. According to \citet{weber2018impact} a climate change induced increase in the seasonal variability of wind speed and a higher likelihood for periods of low wind speeds will lead to significantly higher needs for backup energy and storage in large parts of Europe. This finding is confirmed by \citet{kozarcanin201921st}. They report an increase in balancing and reserve needs of up to 5~\%. \citet{peter2019does2} showed that anticipating climate change in power system planning can reduce system costs in 2100. Accordingly, applying an anticipating strategy increases the optimal share of offshore wind power and decreases the share of nuclear, onshore wind and solar PV power.

Aim of this paper is the description of the sensitivity of a common power system expansion model to different model parameters and designs. We quantify this sensitivity via a novel misallocation metric. This metric measures the additional cost arising from misallocating generation capacities. It allows to cross-validate input data and find, for instance, representative data sets. The method is tested on four different test cases all revolving around a future highly renewable European power system.


\section{Defining a misallocation metric}
Consider a linear program of the form:
\begin{equation}
    \begin{aligned}
        &&\min_x \quad &c^T x \\
        &\text{s.t}& &Ax \geq b \\
        &&&x \geq 0
    \end{aligned}
\end{equation}

In order to solve this program, the parameter matrix $A$, the parameter vector $b$ and the objective coefficients $c$ need to be defined. They determine the actual problem. If $A$ and/or $b$ and/or $c$ are modified, the solution of the linear program changes depending on the sensitivity to the respective parameters.

Now, let $x^*_{\alpha_i}$ be a realisation of the random variables $\{x_1, x_2, \dots, x_{|x|}\}$ solving the linear program under a given set of parameters (in the following referred to as \emph{scenario}) $\alpha_i$. Furthermore, let $\Gamma^{\alpha_i}_0$ be the optimal (minimum) value of the objective function, i.e.
\begin{equation}
\Gamma^{\alpha}_0 = c^T x^*_{\alpha_i}
\end{equation}
If one is interested in the effect of using another different set of parameters expressed as scenario $\alpha_j$, one could, for instance, measure the difference in the objective function value $\Gamma^{\alpha_i}_0 - \Gamma^{\alpha_j}_0$ or the Euclidean norm of the optimal values of the decision variables, $\lVert x^*_{\alpha_i} - x^*_{\alpha_j} \rVert$. The difference between the two solutions can then be distinguished into four cases:


\begin{enumerate}
    \item small difference in the objective function value and \label{en:small_OQM}
        \begin{enumerate}
            \item small difference in the optimal realisation of the decision variables: In this case, the error is obviously small. The linear problem can be considered \emph{insensitive} to the choice of the two scenarios considered. \label{en:subcase_complicated_1}
            \item large difference in the optimal realisation of the decision variables: Here, drawing a clear conclusion is difficult. On the one hand, the error when deciding for one solution might indeed be large. Then, the difference in the objective function value would be small only by chance and we could consider the linear problem to be \emph{sensitive} to the choice of the scenario. On the other hand, the error might be small because the solution space is flat near the optimal point or a secondary optimum exists. The linear problem, again, is insensitive.
        \end{enumerate}
    \item large difference in the objective function value and
        \begin{enumerate}
            \item small difference in the optimal realisation of the decision variables: This case is possible if the linear program is very sensitive towards changes in some decision variables or if the parameters strongly differ between both scenarios. We consider the linear problem to be insensitive.
            \item large difference in the optimal realisation of the decision variables: In this case, the error would obviously be large. The linear problem is sensitive towards the input dataset. \label{en:subcase_complicated_2}
        \end{enumerate}
\end{enumerate}
In order to determine the actual error originating from considering only one of the two scenarios (e.g. $\alpha_i$), one could set the solution $x^*_{\alpha_i}$ as lower bound to the respective other linear program from scenario $\alpha_j$ and vice versa. If the program is insensitive, this should not cause any large additional cost. However, if the program is sensitive, this should cause large additional cost because large adaptations to $x^*_{\alpha_i}$ are necessary in order to make it a solution of the linear program from scenario $\alpha_j$. Let us denote the optimal value of the objective function with lower bounds defined by the optimal solution of the linear program  $\Gamma^{\alpha_j}_{\alpha_i}$ and $\Gamma^{\alpha_i}_{\alpha_j}$, respectively. The additional cost, caused by constraining the solution downwards is then given by $\Gamma^{\alpha_i}_{\alpha_j} - \Gamma^{\alpha_i}_0$ and
the overall sensitivity of the linear program to the choice of the scenario can be quantified by the following \emph{misallocation metric}:

\begin{align}\label{eq:MQM}
\mathrm{M}^{\alpha_i}_{\alpha_j} &= \Gamma^{\alpha_i}_{\alpha_j} - \Gamma^{\alpha_i}_{0} + \Gamma^{\alpha_j}_{\alpha_i} - \Gamma^{\alpha_j}_{0} 
\end{align}



As stated above, $x_{\alpha_i}^*$ refers to the solution of the problem with minimum lower bounds, i.e. the unconstrained problem. $\Gamma^{\alpha_i}_0$ denotes the corresponding value of the objective function. In order to compute the constrained solution $x^*_{\alpha_i,\alpha_j}$ and $\Gamma^{\alpha_i}_{\alpha_j}$, the solution of the corresponding counter-scenario is set as lower bounds to at least some of the decision variables, which means that the following constraint is added to the linear program:
\begin{equation}
 x_i \ge x^*_{\alpha_j}, x_i \in \bar{x} \subset x
\end{equation}
In this paper, these additional constraints are applied to variables representing long-term investment decisions, i.e. generation, storage and transmission capacities.


$M$ is driven by the difference in $\Gamma^{\alpha_i}_0$ and $\Gamma^{\alpha_i}_{\alpha_j}$. When this difference is small, the sensitivity is also small. $M$ fulfills the properties of a pseudometric, it is positive definite
\begin{equation} \label{eq:nonnegativity}
    \mathrm{M}^{\alpha_i}_{\alpha_j} \geq 0
\end{equation}
because $\Gamma^{\alpha_i}_0 \le \Gamma^{\alpha_i}_{\alpha_j}$, symmetric to the order of the scenarios
\begin{equation}\label{eq:symmetry}
    \mathrm{M}^{\alpha_i}_{\alpha_j} = \mathrm{M}^{\alpha_j}_{\alpha_i}
\end{equation}
and fulfills the triangle inequality
\begin{equation}\label{eq:triangle}
    \mathrm{M}^{\alpha_i}_{\alpha_j} \leq \mathrm{M}^{\alpha_i}_{\alpha_k} + \mathrm{M}^{\alpha_k}_{\alpha_j}.
\end{equation}


\section{Methodology}\label{sec:methods}
\subsection{Power System Model and Data}

\begin{table*}[!ht]
    \centering
    \caption{Nomenclature}
    \begin{tabular}{lp{.95\textwidth}}
        \hline \\
         & \\
        $n$, $s$, $t$, $l$ & indices for node, generation/storage type, time and transmission link  \\
         & \\
        $c_{n,s}$ & investment costs for carrier $s$ at node $n$ [EUR/MW] \\
        $\text{CAP}_{\text{CO}_2}$ & global limit on $\text{CO}_2$ emissions [tons] \\
        $\text{CAP}_F$ & global limit of the sum of all single transmission line capacities [MWkm] \\
        $c_l$ & investment costs of transmission capacities at link $l$ [EUR/MWkm] \\
        $d_{n,t}$ & demand at node $n$ and time $t$ [MWh] \\
        $e_{n,s}$ & $\text{CO}_2$ emissions of generators of technology $s$ at node $n$ [tons/MWh] \\
        $\eta_{0,s}$ & standing losses of storage units of technology $s$ [a.u.] \\
        $\eta_{n,s}$ & efficiencies of generators of technology $s$ at node $n$ [a.u.] \\
        $\tau_{n.s}$ & energy-to-power ratio of storage units of technology $s$ at node $n$ [hours] \\
        $\lambda$ & dual variables \\
        $F_l$ & transmission capacities of link $l$ [MW] \\
        $f_{l,t}$ & flows over link $l$ at time $t$ [MWh] \\
        $G_{n,s}$ & capacity of generators or storage units of technology $s$ at node $n$ [MW] \\
        $g_{n,s,t}$ & dispatch of generators or storage units of technology $s$ at node $n$ and time $t$ [MWh] \\
        $g^-_{n,s,t}$ & maximal power uptake of generators or storage units of technology $s$ at node $n$ and time $t$ in units of $G_{n,s}$, zero for generators, negative for storage units \\
        $\bar{g}_{n,s,t}$ & maximum power output of generators or storage units of technology $s$ at node $n$ and time $t$ in units of $G_{n,s}$ \\
        $K_{n,l}$ & incidence matrix of the network \\
        $L_l$ & length of link $l$ [km] \\
        $o_{n,s}$ & marginal costs of generation of technology $s$ at node $n$ [EUR/MWh] \\
        $soc_{n,s,t}$ & state of charge of storage of technology $s$ at node $n$ and time $t$ \\
         & \\
        \hline
    \end{tabular}
    \label{tab:nomenclature}
\end{table*}

\newpage
In this study, we investigate the sensitivity of a common power system expansion problem to (i) different capital cost assumptions (ii) the capacity factor time series for the available volatile renewable resources, (iii) different temporal and spatial resolutions as well as (iv) different model formulations for coupled and decoupled representative periods of different lengths.

The PSEM has the aim to find the least expensive design of a power system given the constraint that the CO$_2$ emissions from power plants may not exceed an upper limit. In the formulation used here, it consists of two levels: While the upper level minimizes the investment cost in generation, storage and transmission capacities keeping all capacities within given bounds, the lower level minimizes the operational cost of the power system ensuring the security of supply and keeping the generation lower or equal the capacities derived from the upper level. For volatile renewable resources -- such as wind and solar PV -- the available dispatchable capacity is additionally limited by prevailing meteorological conditions, i.e. the capacity factor time series. Furthermore, the consistency of the state of charge of storage units must be ensured. Mathematically, the expansion problem can be formulated as a linear problem. For this study, we use the same approach as for instance \citet{brown2018synergies, schlachtberger2017benefits, schlott2018impact}. The linear program is formulated as follows:

\begin{align}
 \min_{g,G,f,F} &\sum_{n,s} c_{n,s} \cdot G_{n,s}  + \sum_{l} c_l \cdot L_l \cdot F_l + \sum_{n,s,t} o_{n,s} \cdot g_{n,s,t} \label{eq:minimisation}\\
\text{s. t.} &\sum_s g_{n,s,t} - d_{n,t} = \sum_l K_{n,l} \cdot f_l \label{eq:powerbalance}\\
&{g}^-_{n,s,t} G_{n,s} \leq g_{n,s,t} \leq \bar{g}_{n,s,t} \cdot G_{n,s} &,\forall n,t \label{eq:dispatchconstraint}\\
&\mathrm{soc}_{n,s,t} = (1-\eta_{n,s}^l) \cdot \mathrm{soc}_{n,s,t-1} + \eta_{n,s}^u \mathrm{uptake}_{n,s,t} &,\forall n, s, t > 1 \label{eq:storage_cont}\\
&\mathrm{soc}_{n,s,0} = \mathrm{soc}_{n,s,|t|} &,\forall n, s \label{eq:storage_cycl}\\
&0 \le \mathrm{soc}_{n,s,t} \le \tau_{n,s} \cdot G_{n,s} \label{eq:soc_bounds}\\
&|f_l\left(t\right)| \leq F_l &,\forall l \label{eq:transm_constraints}\\
&\sum_l F_l \cdot L_l \leq \mathrm{CAP}_{F} \label{eq:global_transm_constr}\\
&\sum_{n,s,t} \frac{1}{\eta_{n,s}} \cdot g_{n,s,t} \cdot e_{n,s} \leq \mathrm{CAP}_{\text{CO}_2} \label{eq:co2cap}
\end{align}

For an explanation of the used symbols see the nomenclature (Tab. \ref{tab:nomenclature}). Constraint (\ref{eq:powerbalance}) describes the balance between generation and demand. Constraints (\ref{eq:dispatchconstraint}) - (\ref{eq:transm_constraints}) effect the dispatch and state of charge of generators, storage and transmission. The dispatch is constrained by the capacity -- or nominal power -- of the respective generator and/or storage unit (\ref{eq:dispatchconstraint}). 
In the case of storage units, the lower bound can be negative, i.e. when the storage takes up energy. In the case of generation technologies, the lower boundary equals zero. The potential generation $\bar{g}_{n,s}(t)$ describes the resource availability in case of fluctuating renewable generation facilities.
Constraints (\ref{eq:storage_cont}) and (\ref{eq:storage_cycl}) ensure storage consistency and cyclic usage of storage, i.e., state of charge at the beginning equals state of charge at the end of the investigated period. Constraint (\ref{eq:soc_bounds}) defines the bounds for the storage unit's state of charge. In equation (\ref{eq:storage_cont}) $\mathrm{uptake}_{n,s,t}$ refers to the net energy uptake of the storage unit given by
\begin{equation*}
    \mathrm{uptake}_{n,s,t} = \eta_1 \cdot g_{n,s,t,\textrm{store}}  -  \eta_2^{-1} \cdot g_{n,s,t,\textrm{dispatch}}
    + \mathrm{inflow}_{n,s,t} - \mathrm{spillage}_{n,s,t}
\end{equation*}
where $\eta_{1,2}$ denote the efficiencies for storing and dispatching electricity, respectively. $g_{n,s,t,\textrm{store}}$ denotes the storing of electricity into the storage unit, $g_{n,s,t,\textrm{dispatch}}$ the dispatch. $\mathrm{inflow}_{n,s,t}$ is the natural inflow into the water reservoir of dams. And $\mathrm{spillage}_{n,s,t}$ denotes the amount of the natural inflow, which is spilled.
In addition, global limits on transmission and CO$_2$ emissions are enforced (Eq. \ref{eq:global_transm_constr} and \ref{eq:co2cap}, respectively). For this paper, we assumed a global limit of three times today's net transfer capacities ($3 \cdot 31.25~\text{TWkm}$) as an appropriate compromise between cost-optimal extension and technical and social concerns. Although, this assumption is slightly more conservative than the compromise grid defined by \citet{schlachtberger2017benefits} and \citet{brown2018synergies} at four times today's values, it allows to capture large parts of the benefits of distributing electricity from renewable resources due to the non-linear decrease in system costs with increasing transmission capacity \cite{schlachtberger2017benefits}. Inline with European emission reduction targets, we define a global CO$_2$ cap of 5\% of the historic level of 1990. In Eq. (\ref{eq:co2cap}), $e_{n,s}$ refers to the emissions given in tonnes of CO$_2$ equivalent per MWh of primary energy. $\eta_{n,s}$ denotes the efficiency of transforming one unit of primary energy into one unit of electrical energy ($g_{n,s,t}$). In this model, OCGT is the only generation type with non-zero CO$_2$ emissions. $e_{n,s}$ and $\eta_{n,s}$ are set to 0.18 tonnes per MWh and 0.39, respectively.

For this study, we use the \emph{PyPSA-EUR} model published by \citet{horsch2018pypsa}. In its full spatial and temporal resolution this model consists of 3567 substations and 6047 transmission lines and covers one year in hourly resolution. It includes time series of capacity factors for onshore and offshore -- where applicable -- wind power as well as solar PV power and time series of electricity demand for each substation. Furthermore, time series for the inflow into hydro reservoirs and runoff river power plants, based on a potential energy approach \cite{kies2016effect}, and upper bounds for the extendable generation capacity per renewable technology and substation are included.

Capacity factor time series are commonly derived from reanalyses data sets \cite{jurasz2020review}. In PyPSA-Eur, time series for wind power capacity factors and the inflow to hydro-electric power plants are derived based on the ERA5 reanalysis \cite{era5}. Onshore and offshore wind power capacity factors have been computed using the power curves of a 3~MW Vestas~V112 with 80~m hub height and the NREL Reference Turbine with 5~MW at 90~m hub height, respectively. Solar PV capacity factor time series have been computed from the Heliosat (SARAH) surface radiation data set \cite{sarah} using the electric model of \citet{huld2010mapping} and the electrical parameters of the crystalline silicon panel fitted in the same publication. All solar panels are assumed to face south at a tilting angle of 35~degrees. Hourly electricity demand for all European countries has been obtained from the \emph{European Network of Transmission System Operators} (ENTSO-E) \cite{entsoedat} and assigned to substations via a linear regression of the GDP and the population. Upper limits of generation capacities  have been derived by restricting the available area to \emph{agricultural areas} and \emph{forest and semi natural areas} given in the \emph{CORINE Land Cover} data set \cite{corine} and by excluding all nature reserves and restricted areas \cite{natura2000}. From the available area, the maximally extendable generation capacity has been computed via fixed densities of 3~MW per square kilometer for onshore wind and 1.45~MW per square kilometer for solar PV, respectively. For further details on the data set and the underlying methodology please see \citet{horsch2017role}.

From this data set, the parameters for the corresponding PSEM (\ref{eq:minimisation})-(\ref{eq:co2cap}) have been defined. Therefore, we fixed the nominal power of all hydro power plants and pumped hydro storage units to the values reported by \citet{kies_restore}, while the nominal power of wind, solar PV and open-cycle gas turbine (OCGT) power plants can be expanded within given bounds. Additionally, we consider two generic storage types with fixed power-to-energy ratio $r$:
\begin{enumerate}
    \item batteries: $r=6$~h
    \item hydrogen storage: $r=168$~h
\end{enumerate}
Their nominal power can be expanded as well. For each technology the investment and operational costs depicted in Tab. \ref{tab:costsassumptions} have been used.

\begin{table}[!ht]
    \begin{center}
        \caption{Cost assumptions for generation and storage technologies, originally based on estimates from \citet{schroder2013current} for the year 2030; fixed operational costs are included in the capital costs.}
        \resizebox{\textwidth}{!}{\begin{tabular}{ lrrrrrr }
        \hline
            Technology & Capital Cost & Marginal cost & Efficiency & Lifetime & energy-to-power ratio \\ 
               &  [EUR/GW/a] & [EUR/MWh] & dispatch/store & [years] & [hours]\\ 
            \hline
            OCGT & 47,235 & 58.385 & 0.390 & 30 &\\
            Onshore Wind & 136,428 & 0.015 & 1 & 25 &\\
            Offshore Wind & 295,041 & 0.020 & 1 & 25 &\\
            PV & 76,486 & 0.010 & 1 & 25 &\\
            Run-Off-River & -- & 0 & 1 & -- &\\
            Hydro Reservoir & -- & 0 & 1 / 1 & -- &\\
            PHS & -- & 0 & 0,866 / 0,866 & -- &\\
            Hydrogen Storage & 195,363 & 0 & 0.580 / 0.750 & 30 & 168\\
            Battery & 120,389  & 0 & 0.900 / 0.900 & 20 & 6\\
            \hline
        \end{tabular}}
        \label{tab:costsassumptions}
    \end{center}
\end{table}

\begin{figure}
    \centering
    \includegraphics[width=.75\textwidth]{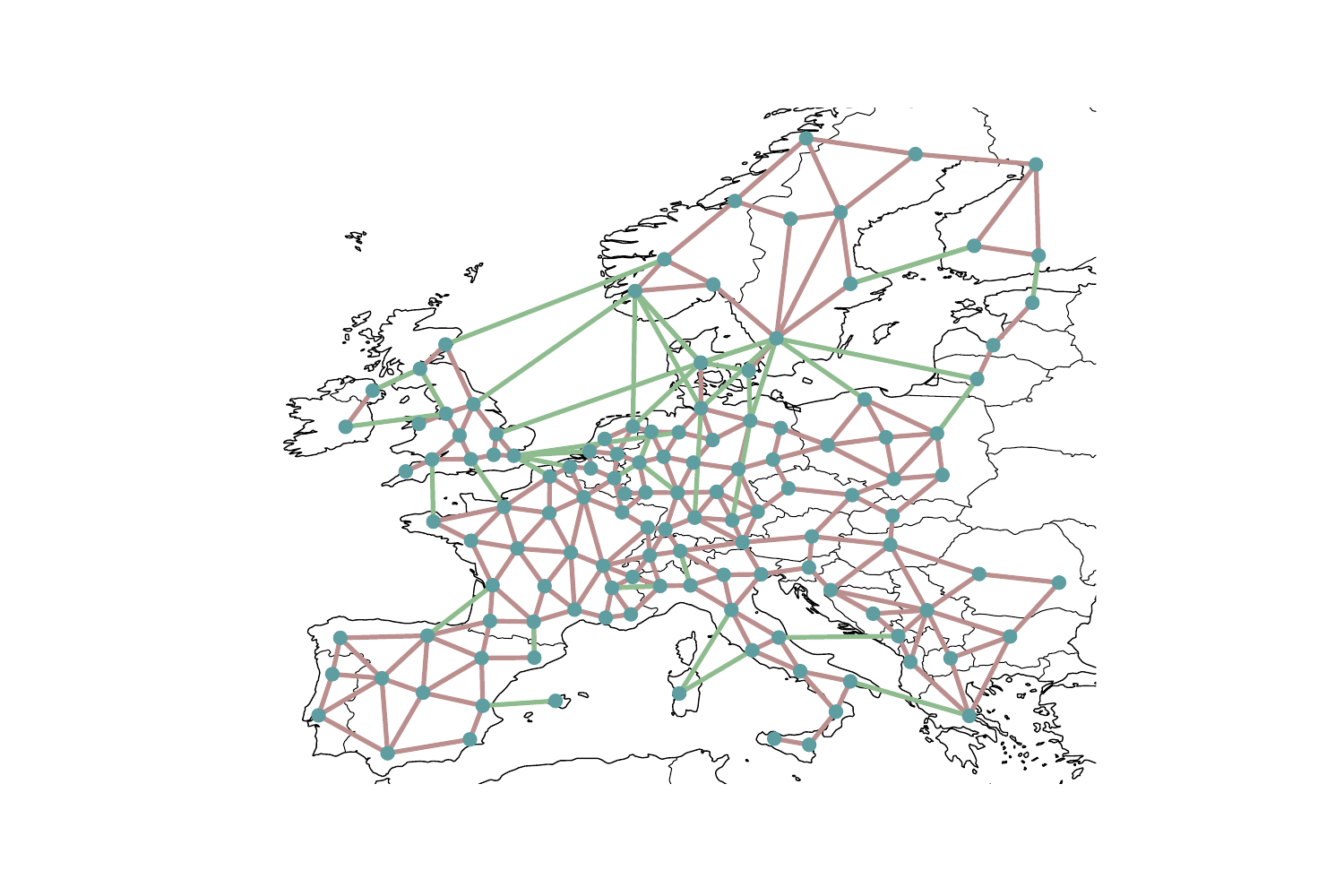}
    \caption{Topology of the PyPSA-Eur network in its 128 node setup.}
    \label{fig:topologies}
\end{figure}

In full resolution, this model can hardly be solved. Therefore, we use the network clustering algorithm introduced by \citet{horsch2017role} to derive clustered versions of the original data set with 45, 64, 90 and 128 substations, respectively. The topology of the network clustered to 128 nodes is shown in Fig. \ref{fig:topologies}. The time series aggregation method described in section \ref{sec:tsam} is then applied to these clustered networks.

\subsection{Reducing the Spatial and Temporal Resolution}\label{sec:tsam}

In general, two types of time series aggregation methods can be distinguished. The first one aims at decreasing the number of time steps by reducing the resolution of the parameter time series. The \emph{downsampling} approach described below, for instance, can be assigned to this class.

The second class aims at decreasing the number of time steps, while keeping the temporal resolution unchanged. In this way, as much of the temporal variability as possible shall be conserved. Usually, this is achieved by selecting a limited number of representative \emph{design periods} from the original time series. Depending on the periods' lengths the variability on different temporal scales can be retained. This, of course, breaks the natural order of the time steps and, consequently, no variability on time scales longer than the periods' lengths can be pictured. Hence, ways need to be found, which allow to model the variability on long time scales (months - seasons), which is represented by the natural inflow into hydro power plants or the seasonal cycle in electricity demand, e.g.. For an overview of these methodologies see \citet{pfenninger2017dealing} and \citet{ kotzur2018impact}.

In order to account for different time step intervals, weightings need to be defined for each time step considered in the expansion problem: first, in the objective ($w_t$ in Eq. (\ref{eq:minimisation})) and second, in the definition of the storage units' state of charge ($\omega_t$ in Eq. (\ref{eq:storage_cont})).

For this study, we applied a simple downsampling technique. It averages the original exogenous parameter time series over consecutive time spans of length $\tau$. Hence, it yields $\nicefrac{T}{\tau}$ time steps at constant intervals. The snapshot weightings $w_t$ and $\omega_t$ are set to $\tau$.


The spatial resolution of the PSEM is modified by applying the network clustering approach introduced by \citet{horsch2017role}. The original model is clustered to 45, 64, 90 and 128 nodes.

\subsection{Computing the misallocation metric}
For each of the parameter sets $\alpha_i$ the expansion problem (\ref{eq:minimisation})-(\ref{eq:co2cap}) is first solved without any lower bounds to the nominal power. The resulting solution vector for the cost-optimal generation capacities $G^*_{n,s}$ is then set as the lower bound to the nominal power for the respective partner problem $\alpha_j$:
\begin{equation}\label{eq:min_constraint}
    \left[G_{n,s}\right]^{\alpha_j}_{\alpha_i} \ge \left[G^*_{n,s}\right]^{\alpha_i}_0
\end{equation}
Following this procedure in both directions delivers the terms of Eq. \ref{eq:MQM}.

In case the number of substations of the two parameter sets differs, i.e. $N_i \not= N_j$, the lower bounds for each parameter set are computed from the corresponding cluster of buses of the other parameter set: Let $\mathcal{N}_i = \{S_{i,1}, S_{i,2}, \dots, S_{i,m}, \dots, S_{i,N_i}\}$, $\mathcal{N}_j = \{S_{j,1}, S_{j,2}, \dots, S_{j,k}, \dots, S_{j,N_j}\}$ be the two sets of clusters of buses derived from the original full-resolution data set with $|\mathcal{N}_{i/j}| = N_{i/j}$. In the clustered networks, each of these clusters $S$ is merged into one single bus $n(S_{i,m})$, $n(S_{j,k})$. Then, the lower bound to a generator of technology $s$ at a bus of set $\mathcal{N}_i$ is set to the weighted sum of the optimal capacity of the buses of set $\mathcal{N}_j$ and vice versa:
\begin{equation}
    \left[G_{n(S_{i,m}),s}^\mathrm{min}\right]^{\alpha_j}_{\alpha_i} = \sum_{k=1}^{N_j} w_k \left[G^*_{n(S_{j,k}),s}\right]^{\alpha_i}_0 \quad ,\forall S_{i,m}, n(S_{i,m})
\end{equation}
where the weights $w_k$ are determined from the number of common nodes of the two clusters $S_{i,m}$, $S_{j,k}$:
\begin{equation}
    w_k = \frac{|S_{i,m} \cap S_{j,k}|}{|S_{j,k}|}
\end{equation}
Here, $|S_{i,m} \cap S_{j,k}|$ is the number of nodes, which appears in both clusters, i.e. the clusters' intersection. 

\section{Results}
\subsection{The Sensitivity to the Capital Costs of Generation Capacities}

\begin{figure}
    \centering
    \includegraphics[width=.48\textwidth]{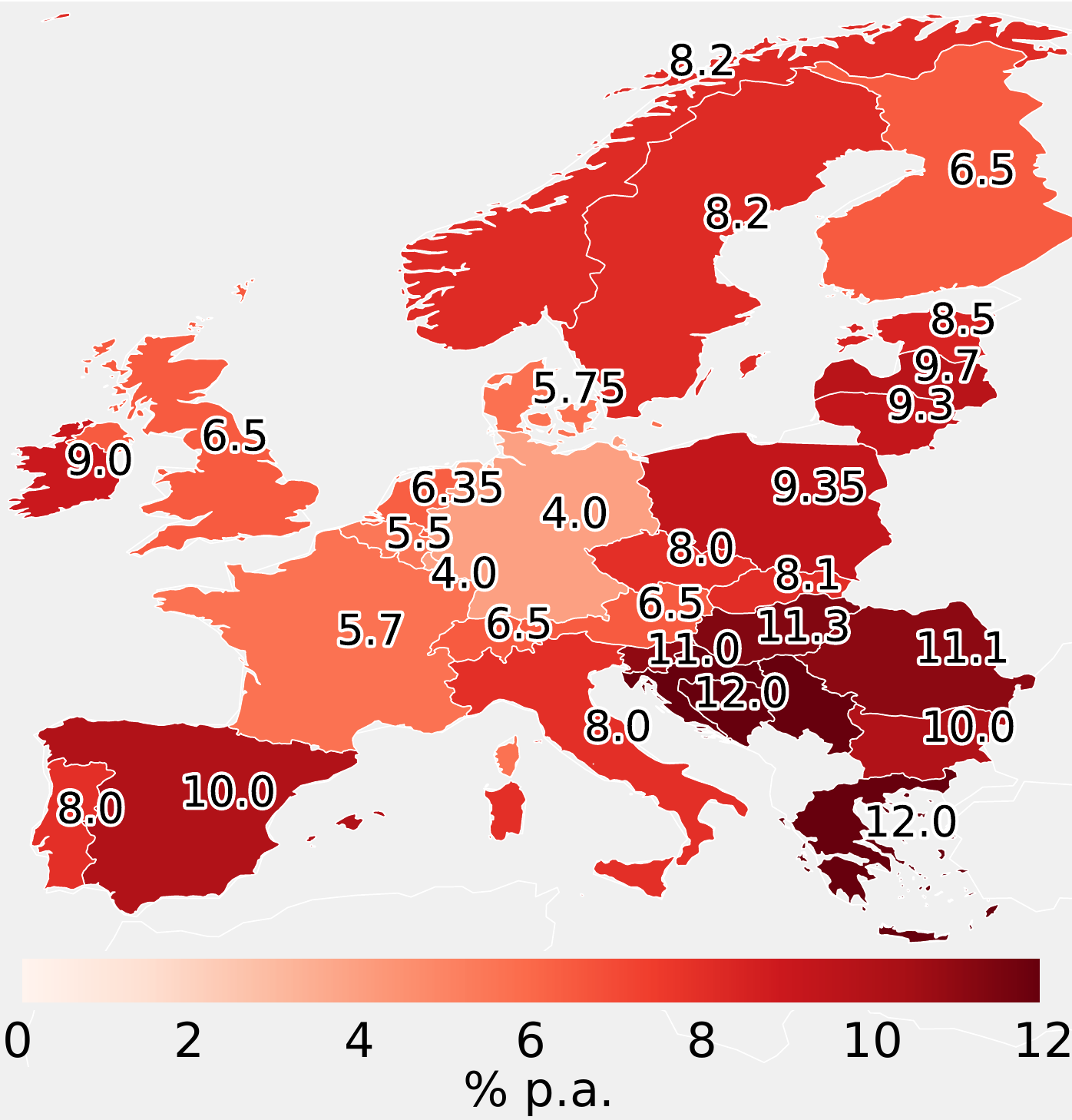}
    \caption{Weighted average cost of capital taken from \citet{noothout2016diacore}, adopted from \citet{schyska2019implications}.}
    \label{fig:wacc}
\end{figure}

As explained above, the capital cost for all generation, storage and transmission assets need to be specified as parameters to the PSEM. Let us, as a first example, assume we would like to investigate the sensitivity of the model to different specifications of these capital cost. In order to do so, we define two scenarios: The first scenario (\emph{hom}) assumes a homogeneous distribution of the cost of capital across the European countries. The second scenario (\emph{dia}) takes regional differences into account. The cost of capital are set as reported from the \emph{diacore} project \cite{noothout2016diacore}. In this scenario, the rate of return on capital ranges from 12\% in the South-Eastern European Countries to only 4\% in Germany (Fig. \ref{fig:wacc}).

\citet{schyska2019implications} have shown that these two scenarios lead to significantly different solutions for the cost-optimal generation capacity layout $x^{*}_{hom}$ and $x^{*}_{dia}$ and the system cost $\Gamma^{hom}_0$ and $\Gamma^{dia}_0$. In particular, the optimal solution for the \emph{dia} scenario contains a larger share of offshore wind power, while the share of onshore wind power, solar PV and OCGT decreases compared to the \emph{hom} scenario. After computing $\Gamma^{hom}_{dia}$ and $\Gamma^{dia}_{hom}$, one finds
\begin{equation*}
    \frac{M^{hom}_{dia}}{\sum_{n,t} d_{n,t}} = 4.6~\frac{\mathrm{EUR}}{\mathrm{MWh}}
\end{equation*}
which is 6.5~\% (6.4~\%) of $\Gamma^{hom}_0$ ($\Gamma^{dia}_0$). Note that the difference in $\Gamma^{hom}_0$ and $\Gamma^{dia}_0$ suggests a smaller error of only $1.4~\nicefrac{EUR}{MWh}$. By applying our new metric we are able to show that the sensitivity of the power system expansion problem to the regional distribution of the cost of capital is indeed much higher (more than 3-fold) than this difference in the levelized costs suggests.

\begin{figure}
    \centering
    \includegraphics[width=.49\textwidth]{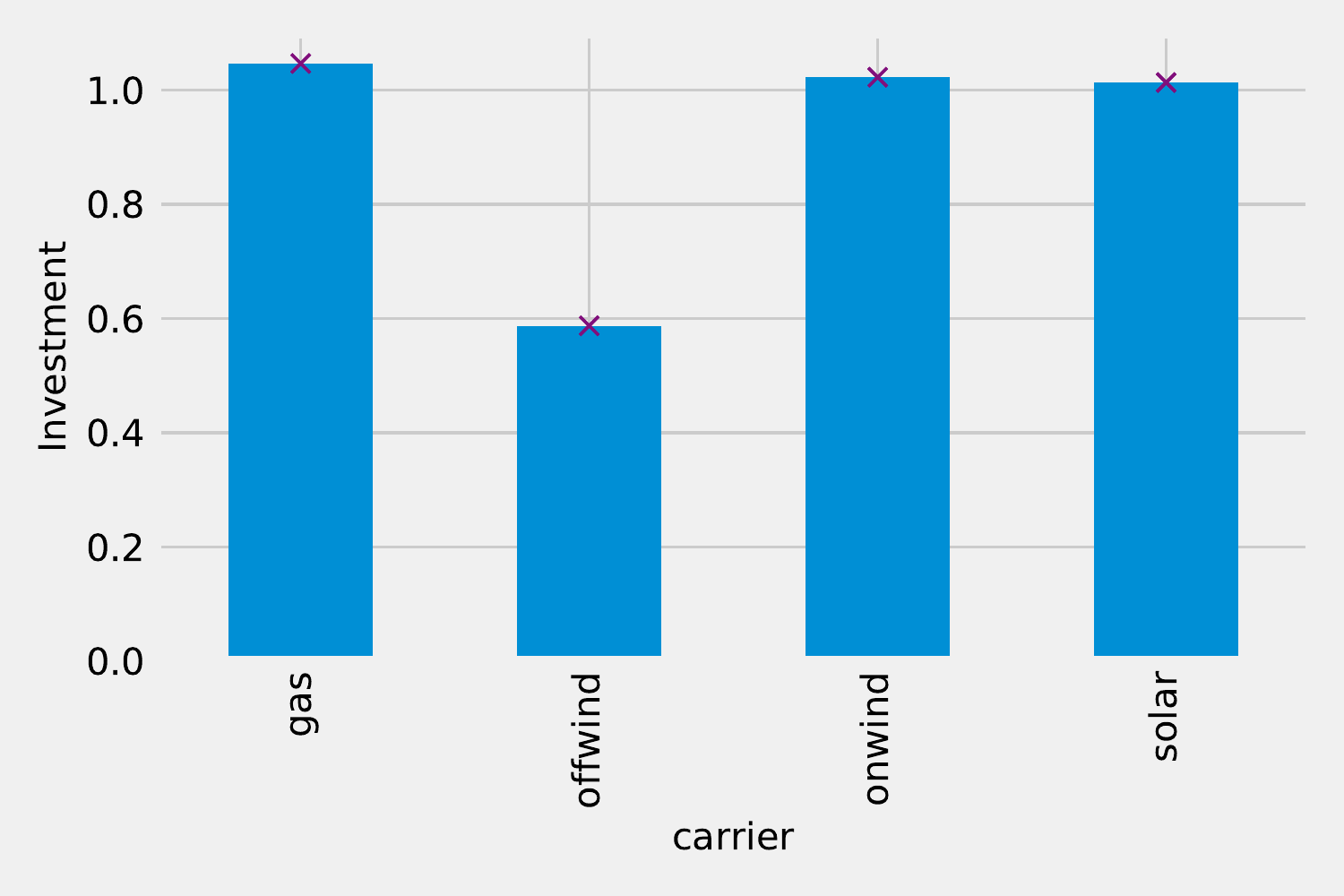}%
    \includegraphics[width=.49\textwidth]{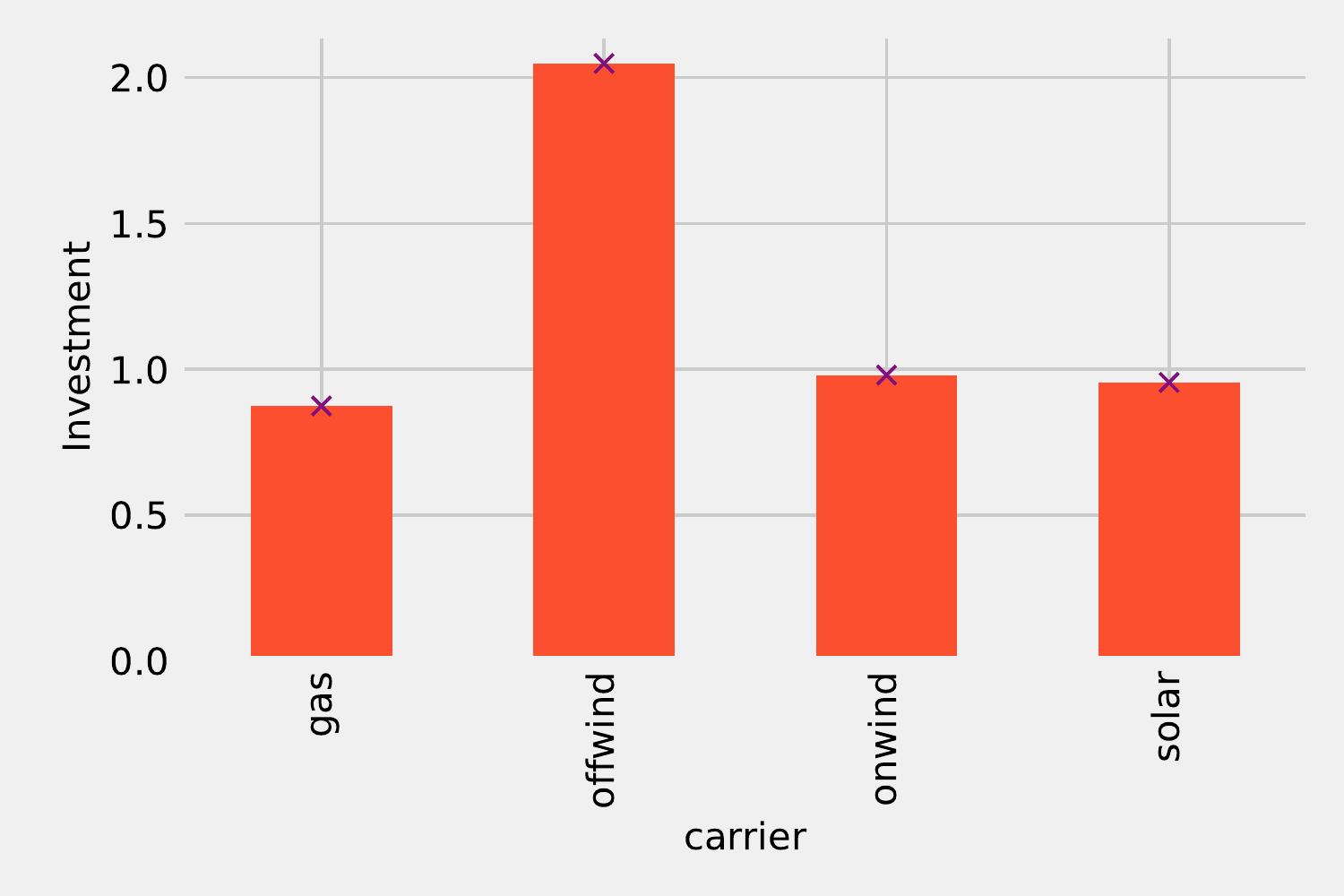}
    \caption{Investment in generation capacity relative to the unconstrained solution for the homogeneous scenario constrained with the solution of the inhomogeneous case (left, blue) and for the inhomogeneous scenario constrained with the solution of homogeneous case (right, orange), crosses indicate the minimum investment for each generation source.}
    \label{fig:wacc_adaptations}
\end{figure}

\begin{figure}
    \centering
    \includegraphics[width=.666\textwidth]{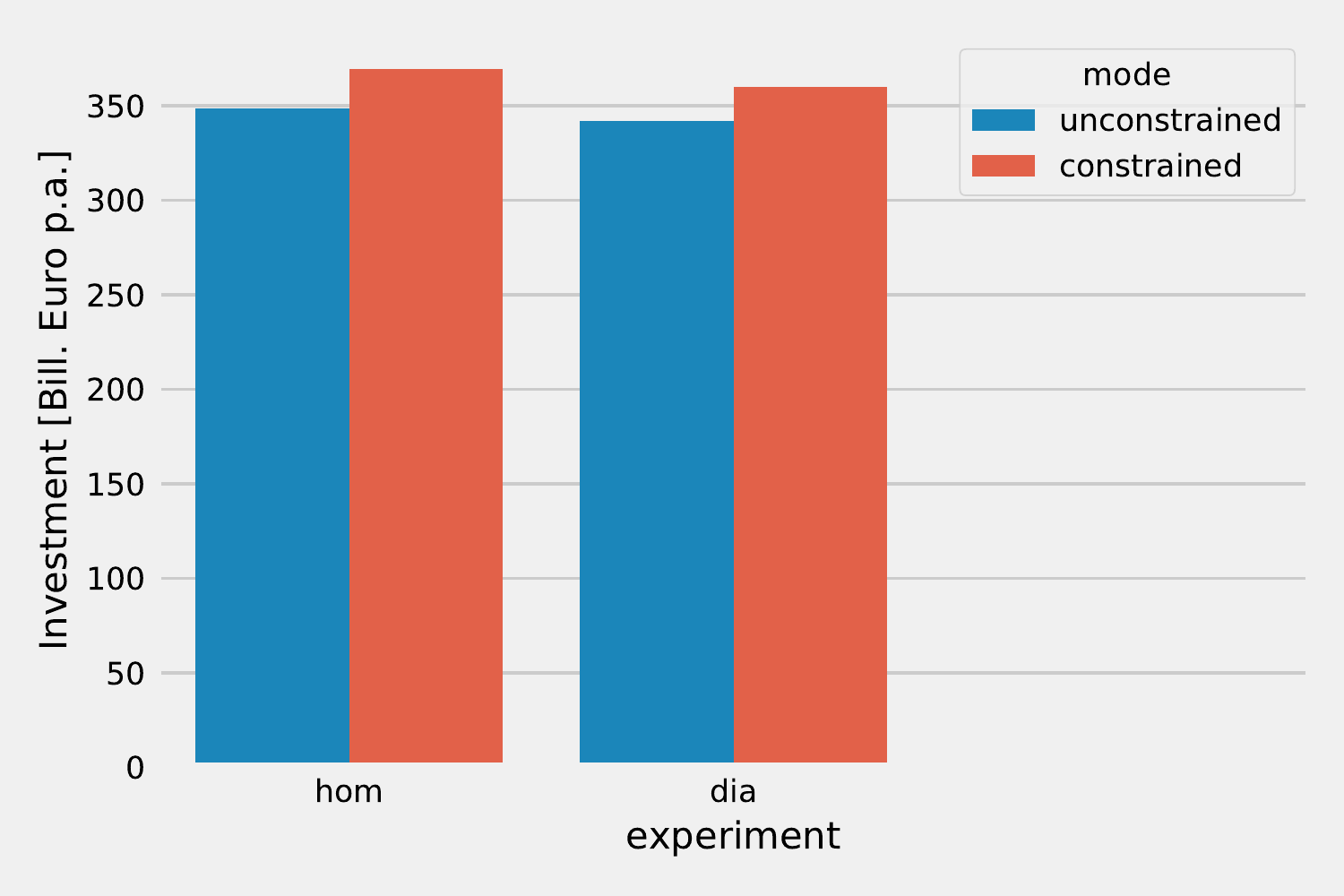}
    \caption{Total investment in generation, storage and transmission capacities in the \emph{hom} scenario (left) and the \emph{dia} scenario (right) and the unconstrained (blue) and constrained case (red) [Bill. Euro].}
    \label{fig:wacc_barplots_investment}
\end{figure}

This higher sensitivity can be explained by taking a look at the shape of the solution space: In general, the solution space for the upper level problem of the expansion problem is steeper than for the lower level problems. This means that slight changes in the capacity layout may lead to significantly different investment cost, while there potentially exist many ways to solve the operational problem with similar cost. This effect is enhanced, if additional regional differences in the cost of capital are considered. Building an offshore wind park in Germany or in Greece, for example, makes a bigger difference now as it made in the homogeneous case. For our first example, we modified the regional distribution of capital cost but kept the nodal loads and the weather time series, which determine the availability of the volatile renewable resources, unchanged. Consequently, we find that both solutions of the unconstrained problems $x^{*}_{dia}$ and $x^{*}_{hom}$ also solve the operational problems of the respective other problem. This is reflected in the fact, that no changes to the capacity layouts are necessary (Fig. \ref{fig:wacc_adaptations}). However, since the two capacity layouts are quite different (as reported above) the investment cost in the constrained cases increase by 6~\% and 5~\%, respectively, compared to the unconstrained cases due to the different cost assumptions in the two scenarios (Fig. \ref{fig:wacc_barplots_investment}).

\subsection{The Sensitivity to the Capacity Factor Time Series}\label{sec:results_capfac}
\begin{figure}
    \centering
    \includegraphics[width=.48\textwidth]{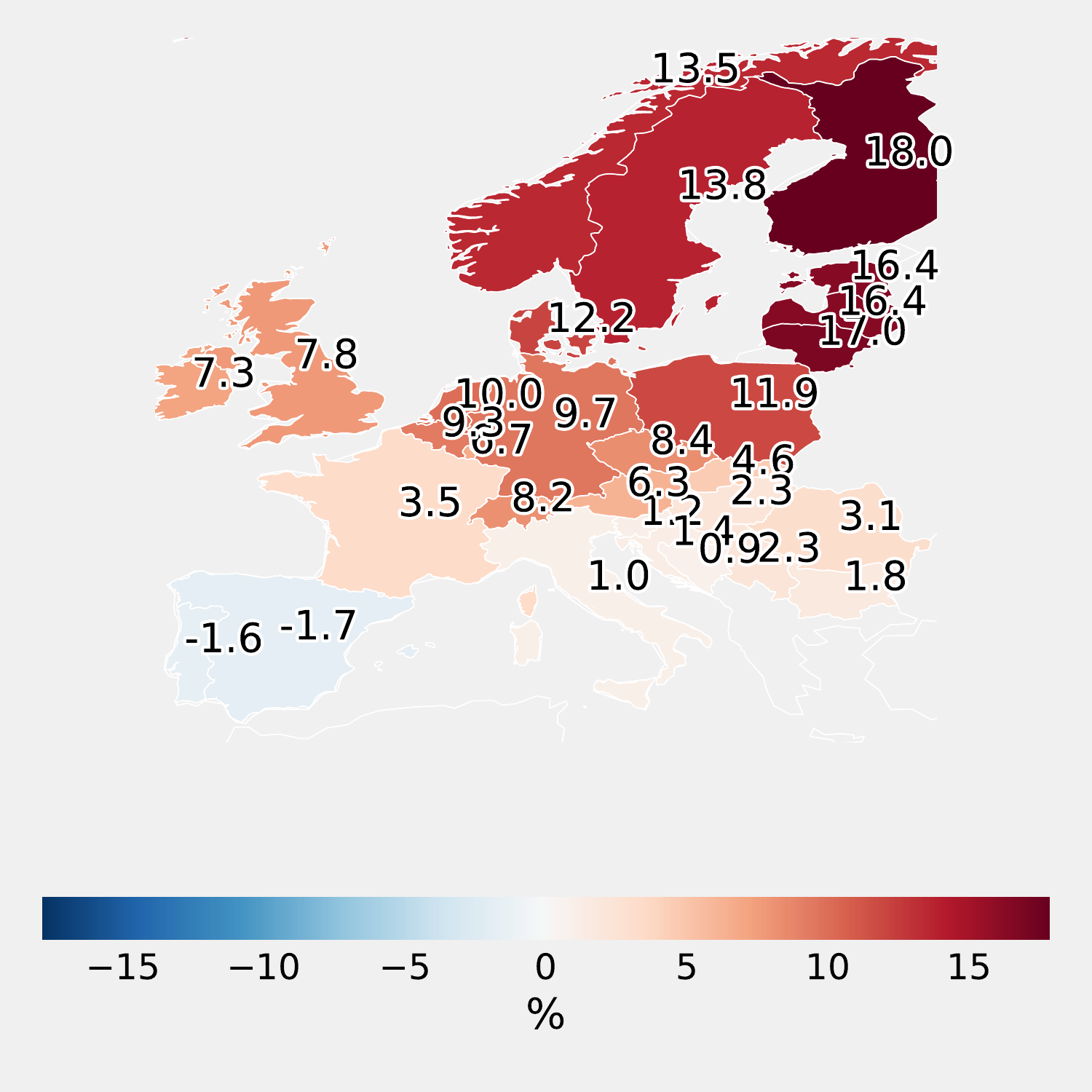}%
    \includegraphics[width=.48\textwidth]{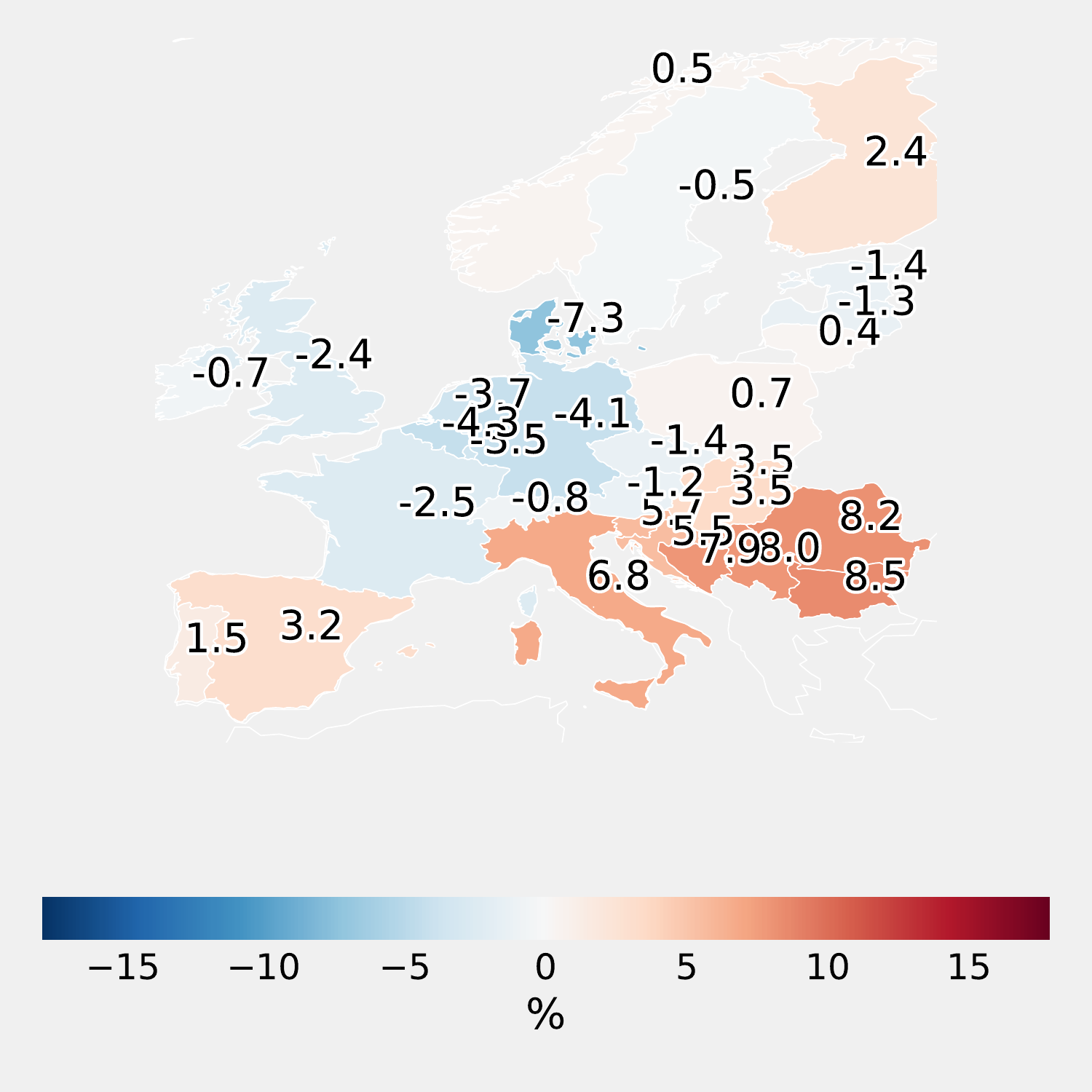}
    \caption{Relative difference in solar power capacity factors (left) and onshore wind power capacity factors (right) between the years 2000-2006 and 2094-2101, [\% of 2000-2006].}
    \label{fig:capfac}
\end{figure}

Second, we investigate the sensitivity of the PyPSA-Eur model to the capacity factor time series $\bar{g}_{n,s,t}$. These time series specify the temporal availability of all volatile resources in the power system at any node. They are given in units of the installed capacity, i.e. $\bar{g}_{n,s,t} \in [0,1]$. For wind and solar power the capacity factor is determined by the prevailing weather situation and, as weather changes from time to time, capacity factors vary as well, from hour to hour but also from year to year and from decade to decade due to climate variability and climate change. Depending on the chosen weather period, the power system optimization might, consequently, lead to different optimal capacity shares. The years 2000-2006, for instance, exhibit a higher average solar power capacity factor over almost entire Europe except for the Iberian Peninsula as predicted for the years 2094-2101. In contrast, the average onshore wind power capacity factor is lower in Central-Western Europe and higher especially in the South-East (Fig. \ref{fig:capfac}). For this investigation, data from \citet{schlott2018impact} has been used. It includes time series of the capacity factors for onshore and offshore wind, solar PV and run-off river as well as of the natural inflow into hydro dams from the climate model \emph{CNRM} \cite{voldoire2013cnrm} downscaled to a higher spatial resolution within the EURO-CORDEX project \cite{jacob2014euro} for each node of the PyPSA-Eur model in its one-node-per-country setup. We use four time slices of length 6-7 years in 3-hourly resolution: 1970-1976, 2000-2006, 2038-2044 and 2094-2101.

\begin{figure}
    \centering
    \includegraphics[width=.95\textwidth]{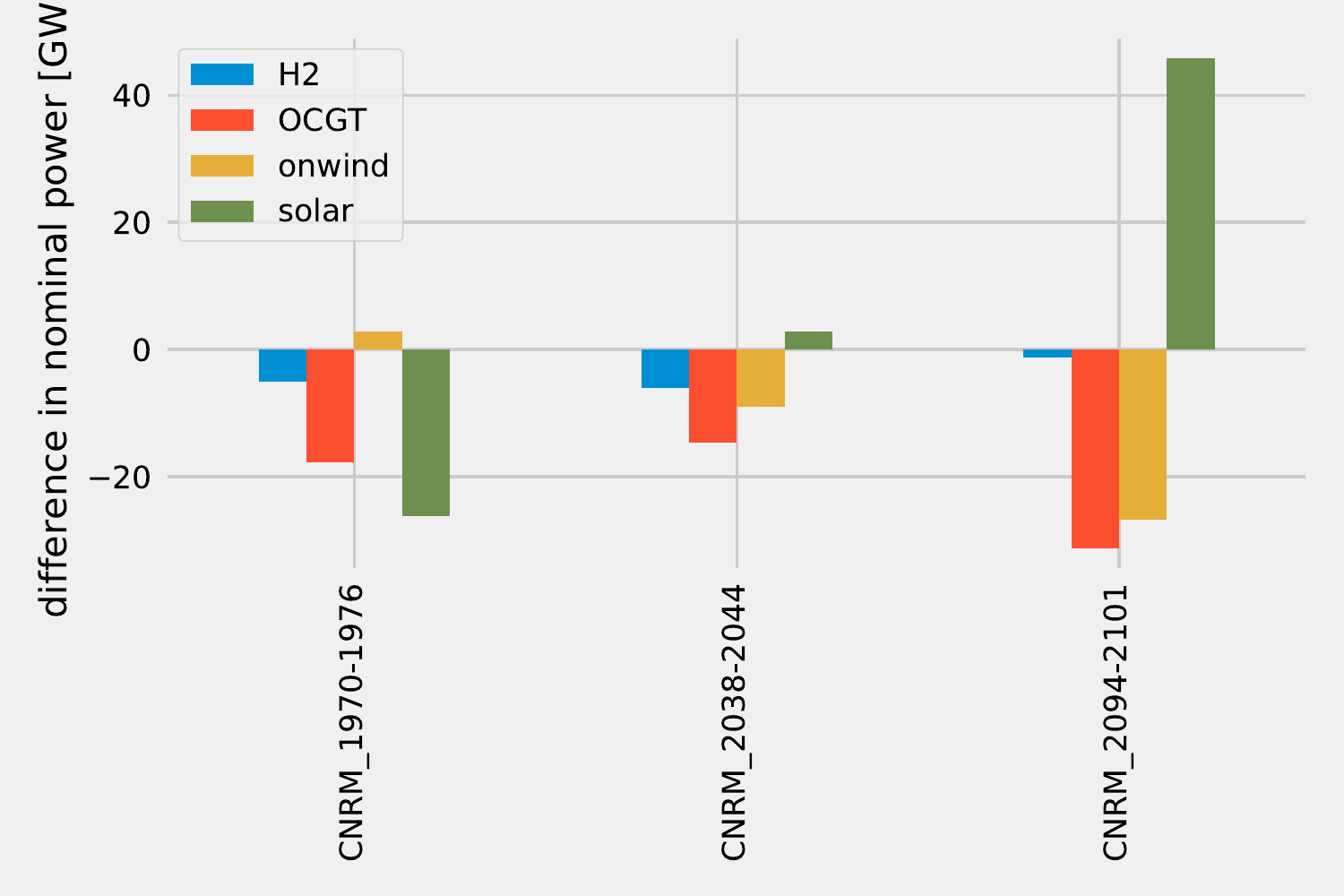}
    \caption{Difference in the cost-optimal capacity deployment compared to the years 2000-2006 [GW].}
    \label{fig:cap_diff_climate}
\end{figure}

The average capacity factor, however, is only one aspect of the quality of renewables resources. The temporal variability, the spatial and temporal co-occurrence of different resources, the correlation with the electricity demand, the system's ability to distribute generation over large areas and the possibility to interact with different kind of storage technologies are equally important. The complex interaction of all these aspects determines the optimal capacity layout of a power system. Indeed, optimizing the PyPSA-Eur model based on either the period 2000-2006 or the period 2094-2101 leads to different optimal investment in generation capacities: Compared to the years 2000-2006, the period 2094-2101 leads to an increase in investment in solar power capacities of 48~\% and a decrease in onshore wind power capacity by 4~\% (Fig. \ref{fig:cap_diff_climate}). Offshore wind is only marginally deployed and can be neglected in both cases.  Interestingly, the levelized cost for all periods differ only slightly. The differences range from 0.16~EUR/MWh for the combination (2038-2044, 2094-2101) to 0.56~EUR/MWh for the combination (1970-1976, 2094-2101). \citet{schlott2018impact} found similar results for CNRM. Other climate models lead to more diverse results in the objective function \cite{schlott2018impact}.

\begin{figure}
    \centering
    \includegraphics[width=.95\textwidth]{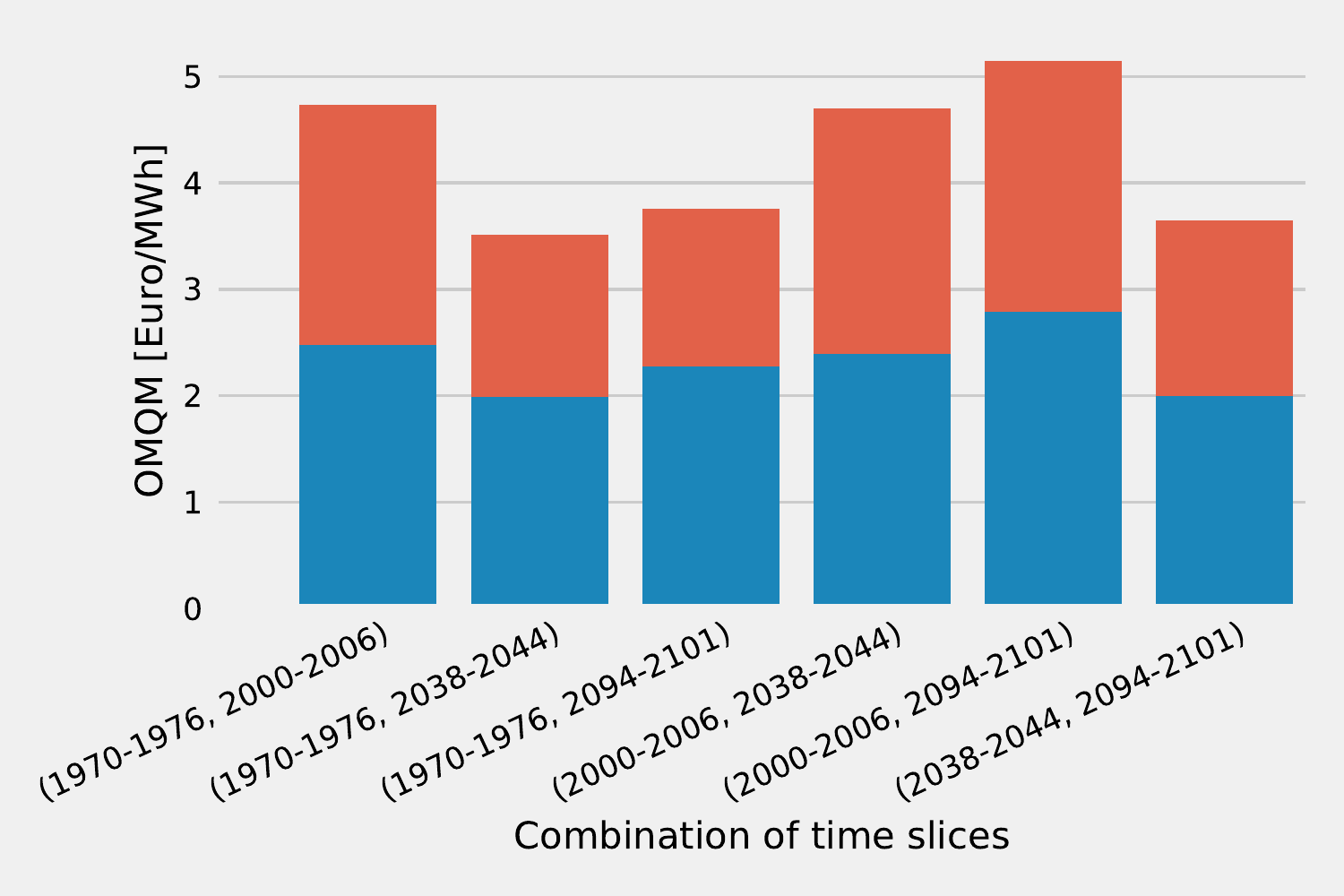}
    \caption{The misallocation metric M [EUR/MWh] for different combinations of weather periods. Blue bars indicate the difference between the constrained and the unconstrained problem of the respective earlier period, the red bar indicates the difference for the later period. Note that for all combinations the difference in the unconstrained solutions is below 0.6~EUR/MWh.}
    \label{fig:heatmap_klimamodelle}
\end{figure}

Let us now compute the sensitivity metric. Therefore, we set the solution for the period 2000-2006 as lower bound to the linear problem with the capacity factors taken from 2094-2101 and vice versa. Adding up the differences in the (two) constrained and unconstrained solutions (Eq. (\ref{eq:MQM})) leads to an overall sensitivity of 5.1~EUR/MWh (Fig. \ref{fig:heatmap_klimamodelle}). This is the highest sensitivity for all possible combinations of the four weather periods considered. The second highest sensitivity is observed for the combinations (1970-1976, 2000-2006) and (2000-2006, 2038-2044). And the combination (1970-1976, 2038-2044) exhibits the lowest sensitivity of 3.5~EUR/MWh. These findings additionally allow to make an inference on the suitability of using a specific climate period for power system investigations. Since the differences between 2000-2006 and the other time periods is the largest, the period 2000-2006 seems not to be representative and hence not a good choice for power system modelling.

\begin{figure}
    \centering
    \includegraphics[width=.49\textwidth]{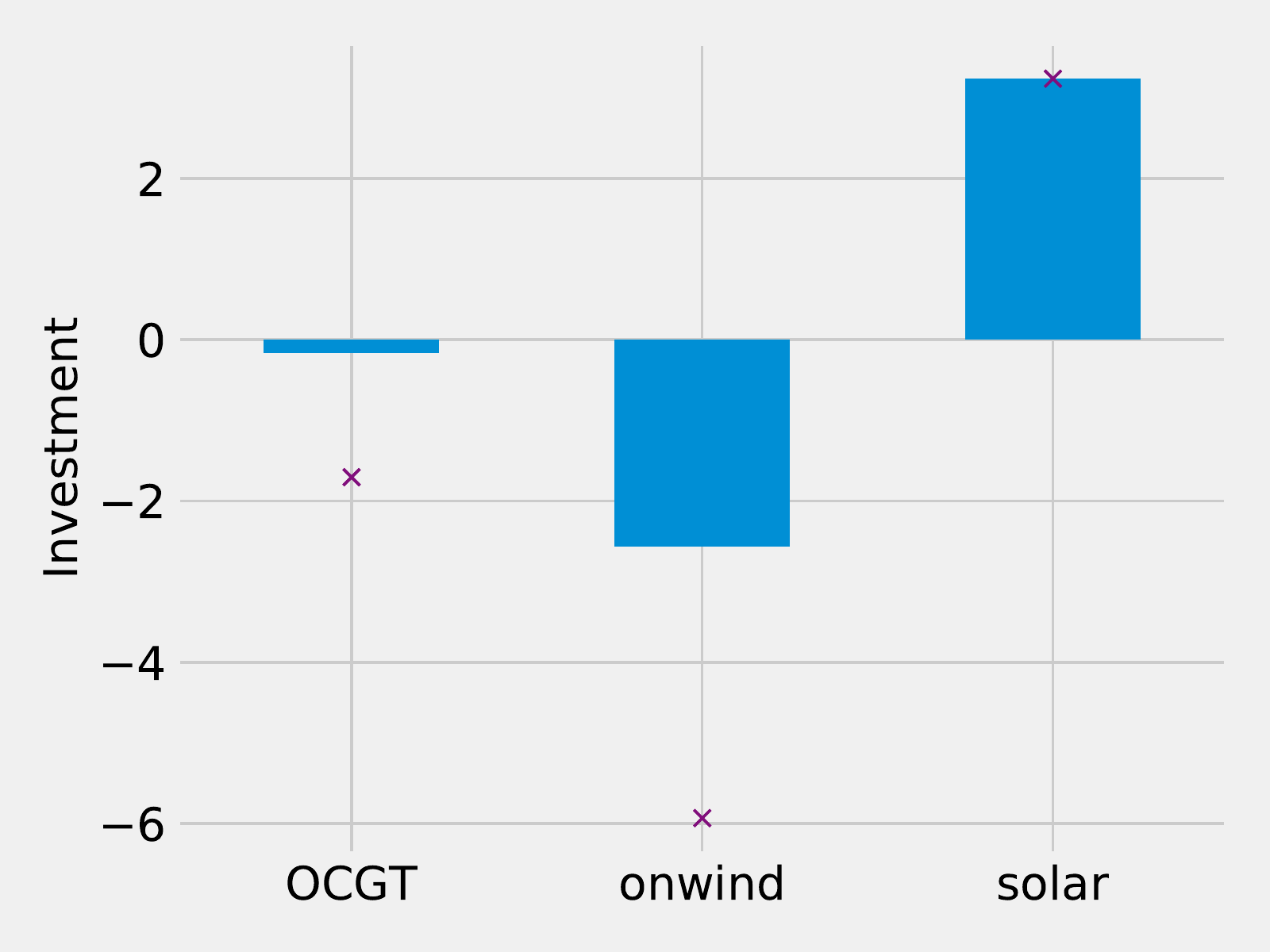}
    \caption{Difference in investment in generation capacity compared to the unconstrained solution for the weather period (2000-2006) constrained with the solution of the weather period (2094-2101). Crosses indicate the minimum investment for each generation source.}
    \label{fig:klimamodelle_adaptations}
\end{figure}

In contrast to the introductory example of modifying the regional distribution of capital cost, we now fixed the cost and varied the availability time series for the different volatile generation sources. In this case, the capacity layout obtained from one optimization is no longer necessarily able to ensure the supply under the constraints of another optimization problem based on another weather period. If this would the case, setting one solution as lower bounds to another scenario would not cause any major additional cost, despite possibly some relatively small changes in the operational cost due to the less effective use of generation, storage and transmission capacities (e.g.). The problem would be insensitive to the differences in the capacity factors from the two scenarios considered, \emph{M} close to zero. However, for the weather periods (2000-2006) and (2094-2101) this is not the case. As mentioned above, optimizing the power system based on (2094-2101) leads to a more solar dominated system, adopted to climate change. Setting this relatively high solar PV capacity as lower bound to the optimization based on (2000-2006) causes major changes in the optimal capacity deployment of the other generation sources as well: The optimal investment in onshore wind power capacity decreases by more than 2.5~Bill. Euro, the investment in OCGT capacity by approximately 0.2~Bill. Euro (Fig. \ref{fig:klimamodelle_adaptations}). This leads to an increase in the \emph{levelized cost of electricity} (LCOE) of 8~\% of the constrained solution compared to the unconstrained solution. This increase in LCOE is the dominating term in Eq. (\ref{eq:MQM}) and determines the sensitivity of the investigated linear problem to the scenarios considered.

\subsection{The Sensitivity to Reduced Spatial and Temporal Resolution}\label{sec:results_base}
\begin{figure}
    \centering
    \includegraphics[width=.95\textwidth]{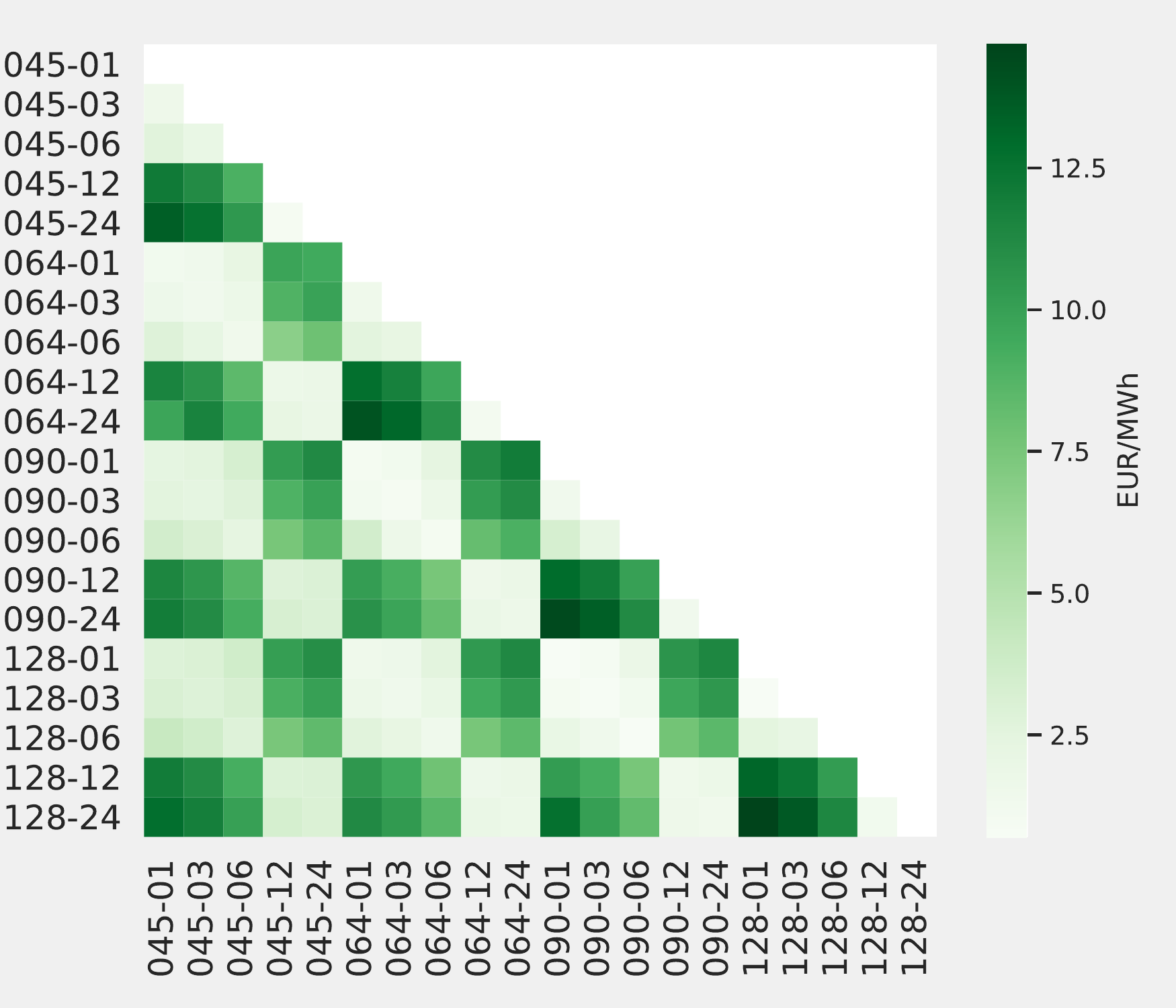}
    \caption{Heatmap of LCOE, normalized to the maximum value [a.u.]}
    \label{fig:heatmap_investment_basemeths}
\end{figure}

Next, we investigate the sensitivity of the linear program (\ref{eq:minimisation}) to the temporal and spatial resolution. In order to reduce the spatial resolution the original PyPSA-Eur network has been scaled down to 45, 64, 90 and 128~nodes using the network clustering approach introduced by \citet{horsch2017role}. The temporal resolution has been reduced by averaging the parameter time series over consecutive time spans of length $\tau \in \{3, 6, 12, 24\}$~hours as described in Sec. \ref{sec:tsam}.

The misallocation metric \emph{M} for all possible combinations of these different parameter sets and model sizes exhibits a clear pattern (Fig. \ref{fig:heatmap_investment_basemeths}). Basically, it can be divided into three different blocks: two blocks of (relatively) low sensitivity where $M \le 4.2$~EUR/MWh and one of (relatively) high sensitivity where $M \ge 7.2$~EUR/MWh. The first block of low sensitivity contains all combinations of scenarios with a temporal resolution higher than 6 hours, i.e. ($N$, 1H), ($N$, 3H) and ($N$, 6H), independent from the spatial resolution $N$. The second block of low sensitivity contains all combination of scenarios with a temporal resolution smaller than 12 hours -- again independent from the spatial resolution. And the block of high sensitivity contains all combination of scenarios where one scenario has high ($\le$ 6 hours) temporal resolution and the other scenario has low temporal resolution ($\ge$ 12 hours). From this definition of blocks one can already see that the expansion problem is much less sensitive to changes in the spatial resolution as it is to changes in the temporal resolution. For instance, the sensitivity of the problem with hourly temporal resolution to increases in the spatial resolution from 45 nodes up to 128 nodes is below 4~EUR/MWh. In contrast, the sensitivity of the 128 node setup to reductions in the temporal resolution from hourly to minimum 12-hourly reaches a maximum value of almost 13~EUR/MWh.

The reasons for this are twofold. First of all, increasing the spatial resolution does not necessarily lead to a higher degree of information in the time series, and vice versa. Consequently, the results obtained from models with different spatial resolutions do not differ much. This phenomenon is of meteorological nature. We denote a separate section to it (Sec. \ref{sec:spatial_vs_temporal}). In contrast, modifying the temporal resolution potentially leads to significant differences in the optimal capacity deployment, especially when the temporal resolution 'jumps' from one of the blocks we defined above to another. Main reason for this is, that downsampling the time series via averaging removes part of the temporal variability. In general, a rolling window averaging can be understood as a filter. For instance, averaging a time series with a rolling 24~hour window filters out most of the sub-24~hour variability of the time series, including the diurnal cycle (if present). If such a filter is applied to both the capacity factor time series and the demand time series, the residual load gets implicitly filtered as well. As a consequence, any storage technology meant to flatten the sub-12~hour variability of the residual load time series would no longer be needed (because there is no sub-12~hour variability).

\begin{figure}
    \centering
    \includegraphics[width=.95\textwidth]{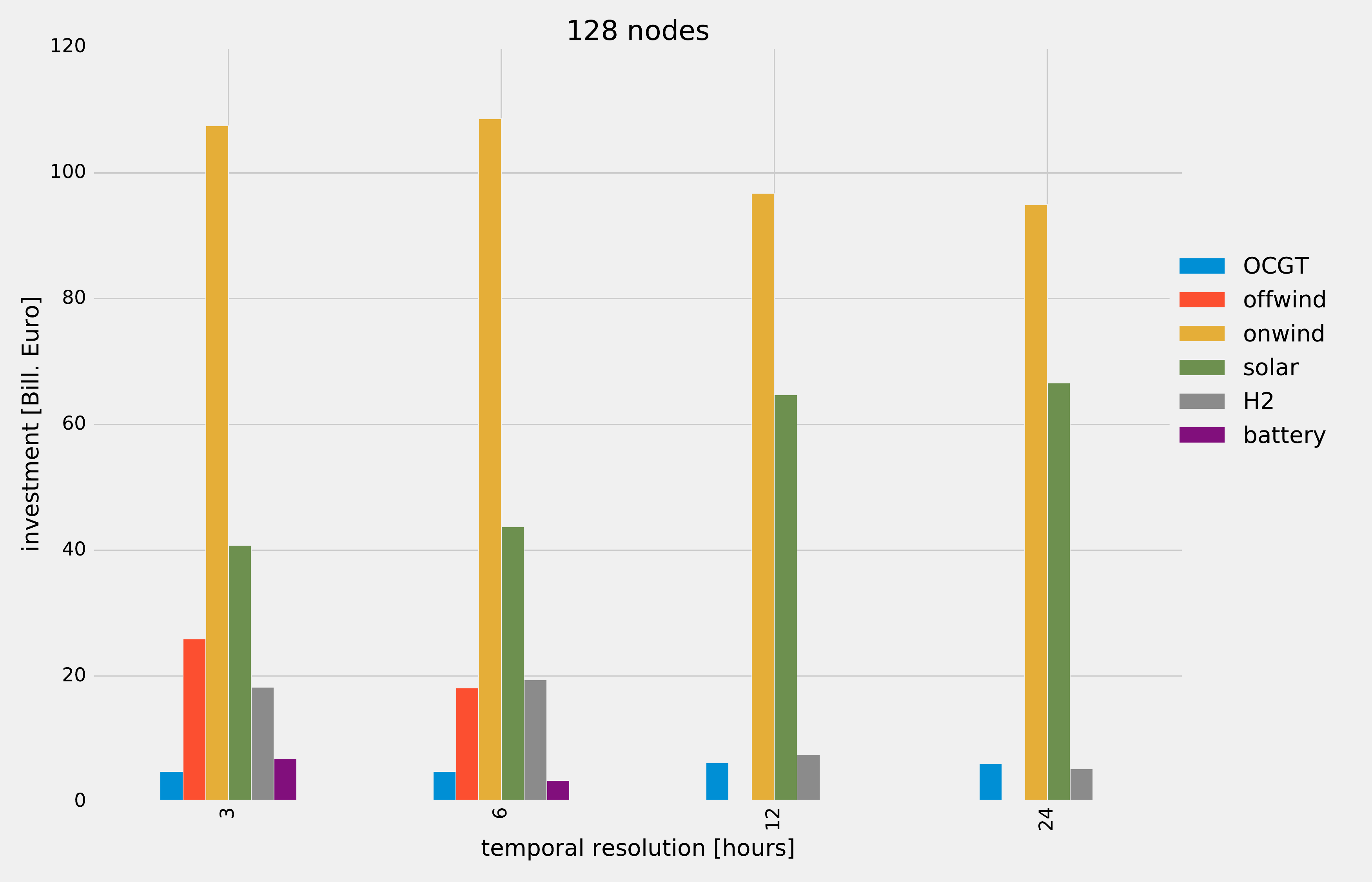}
    \caption{Optimal investment in generation and storage capacity [Bill. Euro] for the 128 node network and for different temporal resolutions of the exogeneous parameter time series.}
    \label{fig:optimal_investment_bars}
\end{figure}

In general, storage technologies can be assigned to a characteristic variability in the residual load time series via their energy-to-power ratio $r$, i.e. the number of hours they can store (dispatch) electricity at full power when starting from empty (full) storage. In the setup used here, batteries are characterized by an energy-to-power ratio of $r=6$~hours. They are meant to balance discrepancies between demand and availability which occur on the intra-day scale. As described above, these discrepancies disappear when the demand and capacity factor time series are downsampled to a lower resolution. Consistently, no battery storage devices are optimally deployed in the model setups with a temporal resolution below 6~hours (Fig. \ref{fig:optimal_investment_bars}). In contrast, hydrogen cavern storage units exhibit an energy-to-power ratio of $r=168$~hours, making them a \emph{weekly} storage. As the weekly variability is still present in the downsampled time series, hydrogen storage devices are still useful.

In broad terms, filtering the high-frequency part of a time series' variability can be understood as removing scatter. This in turn also increases the correlation between the time series, again not only between the availability time series but also between the availability and the demand. Apparently, this rise in correlation mainly increases the system-friendliness of solar PV. Its investment share grows from approximately 40~Bill. Euro for the 3-hourly time series to more than 60~Bill. Euro for the 24-hourly time series (Fig. \ref{fig:optimal_investment_bars}). In turn, the importance of offshore wind power, which is mainly used to cover the baseload in the highly resolved model, decreases, because the filtered time series no longer contain any non-baseload part. The offshore wind power share drops from approximately 23~Bill. Euro to zero. Overall, downsampling time series leads to reduced cost and a significantly different capacity mix. Setting this capacity mix as lower bound to the highly resolved model causes large additional costs, mainly because the model is forced to deploy much more solar PV as it would optimally deploy. Vice versa the offshore and battery storage investment exceeds its optimal value. Overall, this is expressed in a high sensitivity.

However, there is one effect counteracting this phenomenon. This effect appears when the spatial resolution is modified in addition to the temporal resolution. In this case, averaging takes not only part in the temporal dimension but also in the spatial dimension. More precisely, models with a higher spatial resolution experience less averaging on the spatial scale than models with a coarser spatial resolution -- assuming that the models' resolutions are in any case below the resolution of the underlying weather data. This potentially leads to higher capacity factors in the highly resolved case. When transmission capacity is sufficiently available and/or the network is sufficiently meshed, higher capacity factors require less generation capacity as the model with lower spatial resolution. Setting these relatively low capacities as lower bounds to the coarser model does not lead to any additional costs because the optimal capacities are above these bounds anyhow. The lower bounds are non-binding. Consequently, the sensitivity is determined by the additional cost arising from setting the optimal capacities of the coarser model as lower bounds to the finer resolved model. Apparently, these additional cost are small compared to the costs arising from modifying the temporal resolution. When the spatial resolution is not modified, both differences in the equation for the sensitivity metric (Eq. \ref{eq:MQM}) are non-zero. This causes the sensitivity between two models of the same spatial but different temporal resolutions, i.e. ($N$, 1H) and ($N$, 24H) to be larger as between two models of different spatial and temporal resolutions ($N$, 1H) and ($M$, 24H).





\subsection{Correlation Lengths of Wind and Solar Power}\label{sec:spatial_vs_temporal}


\begin{figure}
    \centering
    \includegraphics[height=4.75cm]{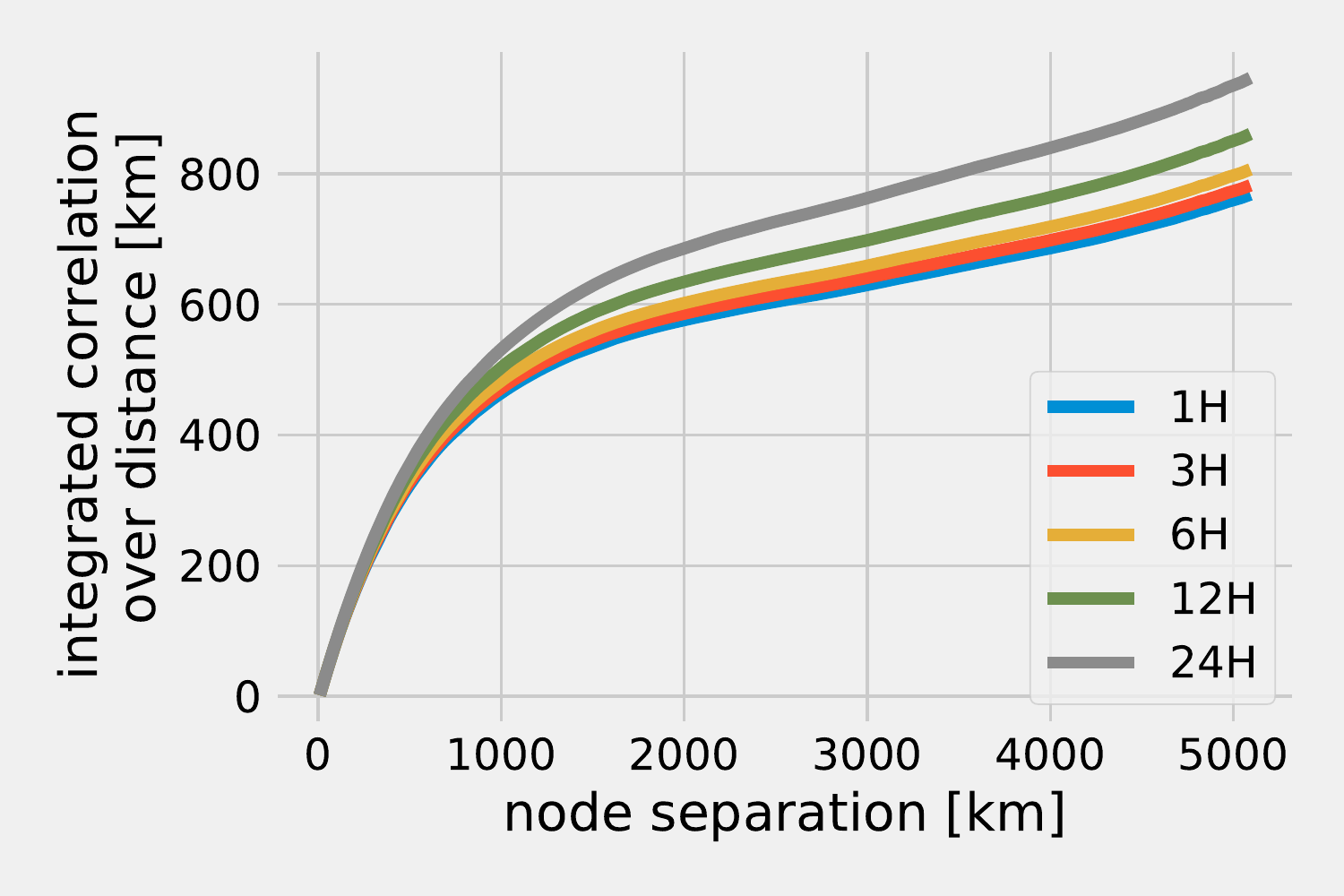}\hskip5pt
    \includegraphics[height=4.75cm]{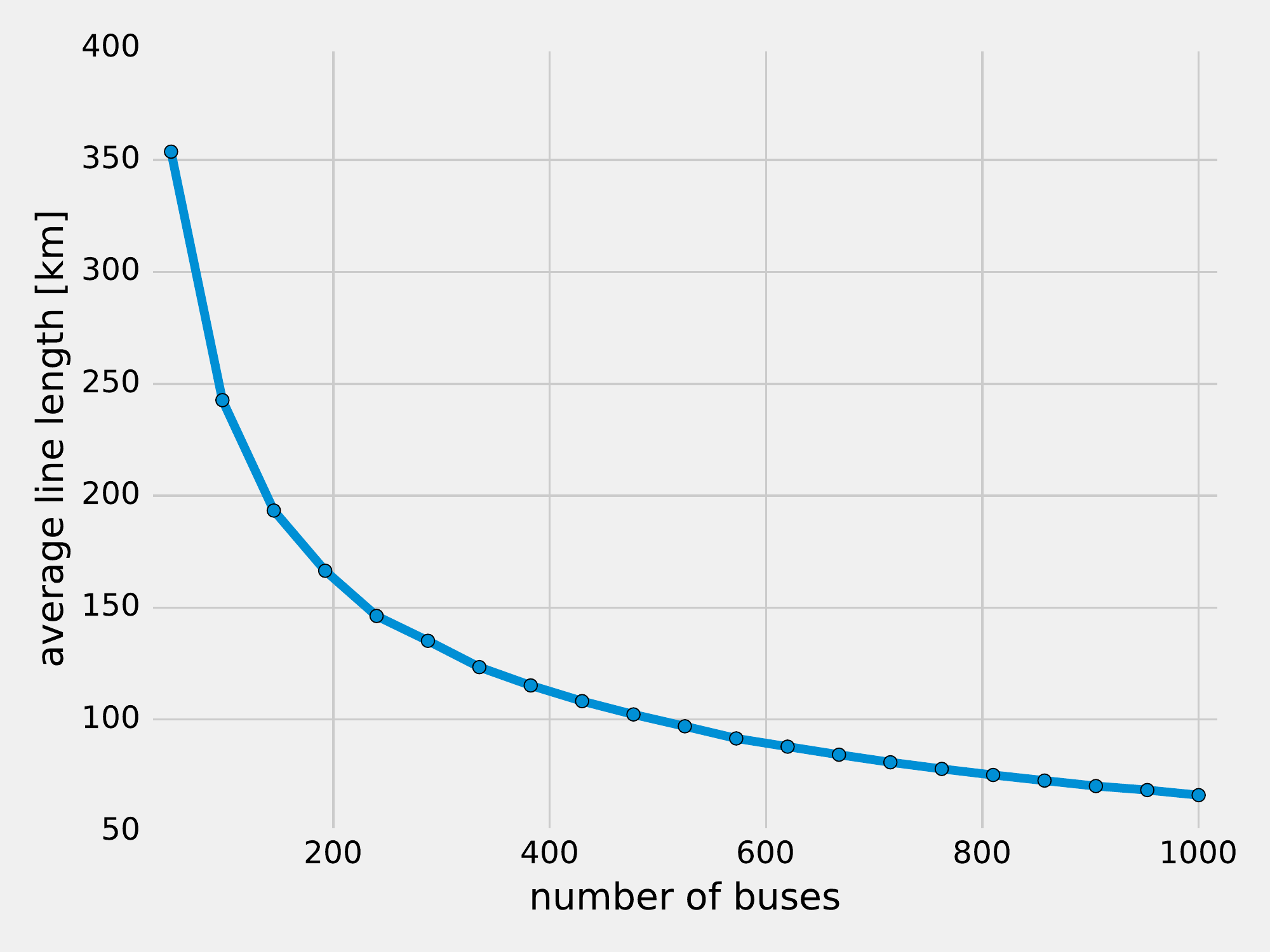}
    \caption{Left: Integrated correlation data of wind power capacity factor time series for the PyPSA-Eur 1024 node setup and different temporal resolutions; the value at largest node separation is referred to as \emph{correlation length}; correlation data is integrated over separation using the trapezoid rule, for further details see \citet{martin2015variability}. Right: Average line length versus number of buses obtained from clustering the original 380-kV network.}
    \label{fig:average_line_length}
\end{figure}

In meteorology, there is a clear relation between a phenomenon's characteristic temporal and spatial scale. Usually, one differentiates between the \emph{microscale}, the \emph{mesoscale} and the \emph{synoptic scale}. While the microscale basically includes turbulent motions acting within second to minutes and with a spatial extent of millimeters to centimeters, the mesoscale covers phenomena like thunderstorms, hurricanes, fronts and convective systems occurring within minutes to days and on several kilometers extent. The synoptic scale includes high and low pressure systems possibly lasting for up to several days on 100 to 1000~kilometers extent \cite{grue2012waves}. Hence, hourly weather time series -- as used here -- only contain the variability introduced by mesoscale and synoptic processes. Sub-hourly microscale processes are filtered out. As mesoscale and synoptic processes act on relatively large spatial scales, the correlation lengths of wind speed can consequently be up to several hundreds kilometers. For wind speed, \citet{martin2015variability} found a correlation length of 273~km in Canada and 368~km in Australia. They computed these correlation lengths for the high frequency, stochastic, part of the time series by applying a high-pass filter and by removing the seasonal cycle prior to estimating the correlation length. However, PSEM have to cope with both aspects of the time series, the high-frequency (stochastic) \emph{and} the low-frequency (deterministic) part. It is the interplay of these two aspects which determines the need for balancing and the optimal capacity share of the respective resource. 
Low frequency variations, in general, exhibit an even higher correlation length as the stochastic time series. For Europe, \citet{schlott2018impact} estimated a correlation length in wind speed of 300 to 700~km which is likely to increase in Northern-Central Europe and to decrease around the Mediterranean towards the end of the century. Without applying any data pre-processing we find that the spatial extent of the power system (5000~km max) is not sufficient to determine the correlation length of the wind power capacity factor by integrating correlation over distance via
\begin{equation}\label{eq:correlation_length}
    \xi(r_n) = \sum_{k=2}^n \frac{1}{2}(r_k - r_{k-1})(\rho_{r_k} + \rho_{r_{k-1}})
\end{equation}
for the PyPSA-Eur 1024~node setup (Fig. \ref{fig:average_line_length} left). For calculating the correlation length, the pairwise distance $r$ and correlation $\rho$ between all nodes have been computed. Correlation data has then been sorted according to node separation. As $\xi(r_n)$ does not saturate until the largest separation, the derived correlation length of approximately 670~km (the integrated correlation at the largest node separation) still is an underestimate of the correlation length which the model experiences.

Compared to wind power, the time series of the solar power availability exhibit a larger deterministic component: the diurnal cycle. On the other hand, incremental changes might be larger. The transition from cloud (shadow) to sun (light) is potentially faster than the transition from windy times to less windy times. 
As sunrise and sunset occur at the same time over large geographical areas, the correlation length of solar power is even larger as the correlation length of wind power.



For the PyPSA-Eur network, the average distance between two nodes varies between 60~km for 1000 nodes and 350~km for the 45 node setup (Fig. \ref{fig:average_line_length} right). These distances are far below the estimated correlation length and although the amount of meteorological information lost depends on the distance of the aggregated nodes, the loss of information -- at least of the kind of information which is relevant for investment decisions -- when buses are aggregated is comparably small. As described above, this is expressed in a relatively small sensitivity to the spatial resolution. Presumably, this would only change when time series with a higher temporal resolution would be used or when an even smaller number of buses would be considered. Time series with a higher temporal resolution would include a higher share of high frequency variability originating from microscale meteorological phenomena. As these phenomena act on smaller spatial scales, the correlation lengths would decrease, too. When the number of buses would be reduced further, the distance between the aggregated nodes might exceed the correlation length. 

Additionally to their temporal variability, time series can be described by  their amplitude. In the context of renewable resource assessment, the quality of the resource is commonly described by the average or the sum of the capacity factor time series, the latter being referred to as \emph{full load hours}. As shown in Sec. \ref{sec:results_base}, this quantity varies as well but on an ever lower frequency, on seasonal to climatological scales. The spatial variance of the full load hours is to a large extent determined by the orography and the latitude. Locations close to the shore, for instance, generally exhibit higher wind power full load hours as locations upcountry. Locations far North are less sunny as locations in the South and, hence, exhibit lower solar power full load hours. Consequently, there is no clear relationship between the distance between two nodes and the difference in the full load hours and the effect of aggregating nodes is hard to assess. It depends on the specific location.


\section{Discussion}
In this study, we introduced a novel method to study the sensitivity of power system optimisation models to different input data scenarios. Core of this method is a metric which is based on setting the decision derived from using one input data set as the lower boundary to the PSEM solving the same program with another parameter scenario. In the sense of modifying and re-solving the original optimization problem it is comparable to the methods applied by \citet{nacken2019integrated} and \citet{neumann2019near}. However, we quantify the sensitivity by one number -- the additional cost arising from misallocating generation, storage and transmission capacities caused by using information for long-term planning which differs from the information the model experiences in short-term operation -- instead of exploring it visually.

In order to test this methodology, we used a relatively simple setup of a European power system model. For instance, we limit the available technologies for electricity generation to OCGT, wind, solar and hydro power. Other technologies such as nuclear or combined-cycle gas turbines are not considered. Furthermore, no coupling of the electricity sector to others sectors is modeled. However, the explanations for the described sensitivities are rather general. We believe that including more technologies and/or incorporating sector coupling would not influence these general findings and the general applicability of the proposed method.

\section{Conclusion}
From the results described above we draw the following conclusions:

\begin{enumerate}
    \item As long as the temporal resolution of the underlying time series does not include any information about microscale meteorological processes, the spatial resolution of the power system model is of minor importance. The sensitivity to increases and decreases in the number of nodes is relatively small. Modeling the European power system with only a few dozens of nodes seems reasonable.
    \item In contrast, the temporal resolution of the underlying time series must be chosen carefully, especially with storage devices involved. The power system model shows the highest sensitivity to modifications of the temporal resolution across the characteristic storage horizon of the storage devices. As a conclusion, the temporal resolution should be chosen such that the variability which the storage devices are supposed to balance is well represented. Particularly, the temporal resolution should be greater than 6-hourly when daily storage units -- such as batteries -- are considered. Contrarily, time series with daily resolution might be appropriate when only weekly and/or seasonal storage types are part of the model.
    \item In summary, we found an on average sensitivity to the choice of the underlying weather data. Our results indicate that the period 2000 through 2006 is not suitable for deriving general conclusions about the optimal design of the European power system. It let to the highest misallocation of generation and storage capacities compared to the other periods considered. This finding emphasizes the importance of using representative weather data sets.
    \item Similarly, the capital cost of generation assets should be defined according to the state of the art. The sensitivity to the geographical distribution of the cost of capital was found to be as high as the sensitivity to the capacity factor time series.
\end{enumerate}

In future research, it seems reasonable to compare and combine the proposed methods with the MGA approach of \citet{nacken2019integrated} or the methods to investigate the shape of the solution space proposed by \citet{neumann2019near} to study the uncertainty of energy system models. Furthermore, the sensitivity to modifications in the temporal resolution could be further investigated by applying the approach of coupling design periods introduced by \citet{gabrielli2018optimal} or the time series aggregation approach based on hierarchical clustering with connectivity published by \citet{pineda2018chronological}.



\section*{References}



\bibliographystyle{model1-num-names.bst}
\bibliography{references.bib}







\end{document}